\documentclass[10pt,conference,letterpaper]{IEEEtran}
\IEEEoverridecommandlockouts
\newfont{\mycrnotice}{ptmr8t at 7pt}	
\newfont{\myconfname}{ptmri8t at 7pt}
\usepackage{latexsym}
\usepackage{epsfig}
\usepackage{subfigure}
\usepackage{amsfonts}
\usepackage{times}
\usepackage{tabularx,amssymb,amsmath,multirow, algorithm, algorithmic,graphicx}
\usepackage{cite}
\usepackage{mdwlist}
\usepackage{blindtext}
\usepackage{verbatim}
\usepackage{url}
\usepackage{balance}
\usepackage{comment}
\usepackage{makecell}
\usepackage{multirow}%\usepackage{silence}
\usepackage[english]{babel}
\usepackage{hhline}  % extended styles for tables
\usepackage{array}
\usepackage{fancyhdr}

\newcommand{\PreserveBackslash}[1]{\let\temp=\\#1\let\\=\temp}
\newcolumntype{C}[1]{>{\PreserveBackslash\centering}p{#1}}
\newcolumntype{R}[1]{>{\PreserveBackslash\raggedleft}p{#1}}
\newcolumntype{L}[1]{>{\PreserveBackslash\raggedright}p{#1}}

\begin{document}

%% Copyright
%\setcopyright{acmcopyright}
%%\setcopyright{acmlicensed}
%%\setcopyright{rightsretained}
%%\setcopyright{usgov}
%%\setcopyright{usgovmixed}
%%\setcopyright{cagov}
%%\setcopyright{cagovmixed}
%
%
%% DOI
%\doi{10.475/123_4}
%
%% ISBN
%\isbn{123-4567-24-567/08/06}
%
%%Conference
%\conferenceinfo{PLDI '13}{June 16--19, 2013, Seattle, WA, USA}
%
%\acmPrice{\$15.00}
%
%%
%% --- Author Metadata here ---
%\conferenceinfo{WOODSTOCK}{'97 El Paso, Texas USA}
%%\CopyrightYear{2007} % Allows default copyright year (20XX) to be over-ridden - IF NEED BE.
%%\crdata{0-12345-67-8/90/01}  % Allows default copyright data (0-89791-88-6/97/05) to be over-ridden - IF NEED BE.
%% --- End of Author Metadata ---

\title{Achieving Spectrum Efficient Communication Under Cross-Technology Interference}
\author{\IEEEauthorblockN{Shuai Wang$^{1}$, Zhimeng Yin$^1$, Song Min Kim$^2$, Tian He$^1$} \\
\IEEEauthorblockA{$^1$Computer Science and Engineering, University of Minnesota, USA \\
$^2$Computer Science, George Mason University, USA \\}}

\maketitle

\begin{abstract}
In wireless communication, heterogeneous technologies such as WiFi, ZigBee and
BlueTooth operate in the same ISM band. With the exponential growth in the number
of wireless devices, the ISM band becomes more and more crowded. These heterogeneous
devices have to compete with each other to access spectrum resources, generating
cross-technology interference (CTI). Since CTI may destroy wireless communication,
this field is facing an urgent and challenging need to investigate spectrum efficiency
under CTI. In this paper, we introduce a novel framework to address this problem
from two aspects. On the one hand, from the perspective of each communication technology
itself, we propose novel channel/link models to capture the channel/link status under CTI.
On the other hand, we investigate spectrum efficiency from the perspective by taking
all heterogeneous technologies as a whole and building cross-technology communication
among them. The capability of direct communication among heterogeneous devices brings
great opportunities to harmoniously sharing the spectrum with collaboration rather
than competition.

\end{abstract}

%
%%
%% The code below should be generated by the tool at
%% http://dl.acm.org/ccs.cfm
%\begin{CCSXML}
%<ccs2012>
%  <concept_id>10010520.10010553.10010562</concept_id>
%  <concept_desc>Computer systems organization~Embedded systems</concept_desc>
%  <concept_significance>500</concept_significance>
%  <concept_id>10010520.10010575.10010755</concept_id>
%  <concept_desc>Computer systems organization~Redundancy</concept_desc>
%  <concept_significance>300</concept_significance>
%  <concept_id>10010520.10010553.10010554</concept_id>
%  <concept_desc>Computer systems organization~Robotics</concept_desc>
%  <concept_significance>100</concept_significance>
%  <concept_id>10003033.10003083.10003095</concept_id>
%  <concept_desc>Networks~Network reliability</concept_desc>
%  <concept_significance>100</concept_significance>
%</ccs2012>
%\end{CCSXML}
%
%\ccsdesc[500]{Computer systems organization~Embedded systems}
%% End generated code
%%
%
%%
%%  Use this command to print the description
%%
%\printccsdesc
%
%% We no longer use \terms command
%%\terms{Theory}
%
%\keywords{ACM proceedings; \LaTeX; text tagging}

\section{Introduction}

%Popular IoT and applications.
%background: dense ISM band;

%the problem; existing solutions;

%Our solutions:

%Three aspects:

Wireless technologies are widely utilized in people's daily life for personal communication, mobile internet surfing,
global positioning, and smart home automation. To accommodate different application requirements on
system performance (e.g., throughput, reliability, delay, and energy consumption), a wide range of wireless technologies,
such as WiFi, BlueTooth and ZigBee, have been proposed. Many of these technologies share the same spectrum, e.g., 2.4G ISM
(industrial, scientific and medical) bands.

%With the popularity of wireless technology, related equipment exponentially increased

With the increasing number of wireless devices, these heterogeneous devices have to compete with each other to access spectrum resources, generating cross-technology interference.
For example, in a residential building, WiFi devices provide wireless internet connectivity for web surfing and video streaming, whereas ZigBee devices enable energy-efficient sensing and actuation for home automation. In close proximity, it has been shown that traffic generated by a WiFi device can disrupt the communication of other WiFi devices or ZigBee devices severely~\cite{beyondcoexistence,SurvivingWiFi,CooperativeCS}.

To alleviate the burden of spectrum shortage and reduce the cross-technology interference, researchers propose dynamic spectrum access~\cite{Akyildiz06nextgeneration,DSAsurvey,Yang2010,Hong2012,Rayanchu} based on cognitive radio to allow different wireless technologies to share the spectrum resources. However, most existing wireless devices (e.g., WiFi, ZigBee, and BlueTooth) have no cognitive capability and the deployment of a commercial cognitive radio network is yet to emerge.

In this paper, we introduce a novel framework for spectrum efficient communication.
The highlight of our work is that the proposed approaches in the framework are compatible
to existing standards (e.g., 802.11 and 802.15.4) and can be implemented directly on
off-the-shelf heterogeneous devices. Our framework studies the spectrum communication
efficiency from two aspects.

On the one hand, we study spectrum efficiency from the perspective of each communication
technology being disturbed by cross-technology interference. We find that existing channel/link models~\cite{MRD,Miluzzo08,multipathg,Sriniva08,Sriniva,shan06,Macro07,Jerry} can
not fully capture the real-world channel/link status under the impact of cross-technology
interference. We thus propose more realistic models which accommodate the temporal and spatial
channel/link dynamics caused by cross-technology interference. Our new link/channel models
have a broad impact on protocol designs including but are not limited to (i) traditional
network protocols such as broadcast~\cite{Lou, Qayyum},
multi-cast~\cite{Murphy, Huang}, and multi-path routing~\cite{Ganesan},  or (ii)
diversity-based protocols such as network coding~\cite{Ahlswede, Katti, lili}, collaborative
forwarding~\cite{Cao} and opportunistic forwarding~\cite{Biswas, Chachulski, Du}.

%We also investigate the spectrum efficiency problem from the perspective of cross-technology
%interference. We propose models to detect, differentiate, and filtrate cross-technology
%interference. These models help us (i) understand the characteristic of interference from WiFi, BlueTooth
%and ZigBee devices, (ii) differentiate the interference source of corrupted symbols, and (iii)
%build corresponding mechanisms to recover the corrected symbols.

On the other hand, we investigate spectrum efficiency by taking all heterogeneous technologies
as a whole and building cross-technology communication among these technologies. The capability
of direct communication among heterogeneous technologies brings us great opportunities to efficiently
utilize the scarce spectrum resources through direct negotiation of spectrum access. In addition, the
realization of cross-technology communication makes us rethink our existing designs on IoT applications
such as device coordination and control in smart home.

The rest of the paper is organized as follows.
Section II introduces the background on spectrum utilization.
Section III reviews related work.
%Section IV and V introduce our framework of achieving spectrum efficient communication.
%:
Section IV introduces new link modeling and its applications under cross-technology interference;
Section V introduces cross-technology communication and its applications.
Finally, Section VI concludes the paper.

\begin{figure}[!htbp]
\centering
\includegraphics[width=0.48\textwidth]{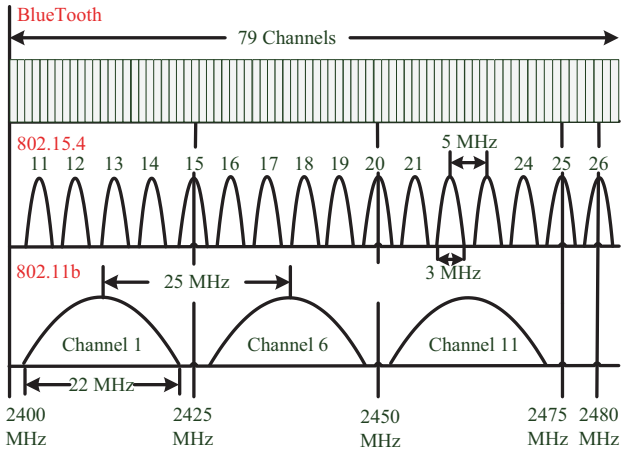}
\caption{The overlapped channels of IEEE 802.11b (WiFi), IEEE 802.15.1 (BlueTooth) and IEEE 802.15.4 (ZigBee)~\cite{srinivasan10tosn}.}
\vspace{-2mm}
\label{ctc:sharespectrum}
\end{figure}

\section{Background}\label{sec:Prel}
\begin{figure}[!htb]
\centering
\subfigure[WiFi\&BlueTooth Devices]{
\label{fig:wifibt}
\includegraphics[width=0.23\textwidth]{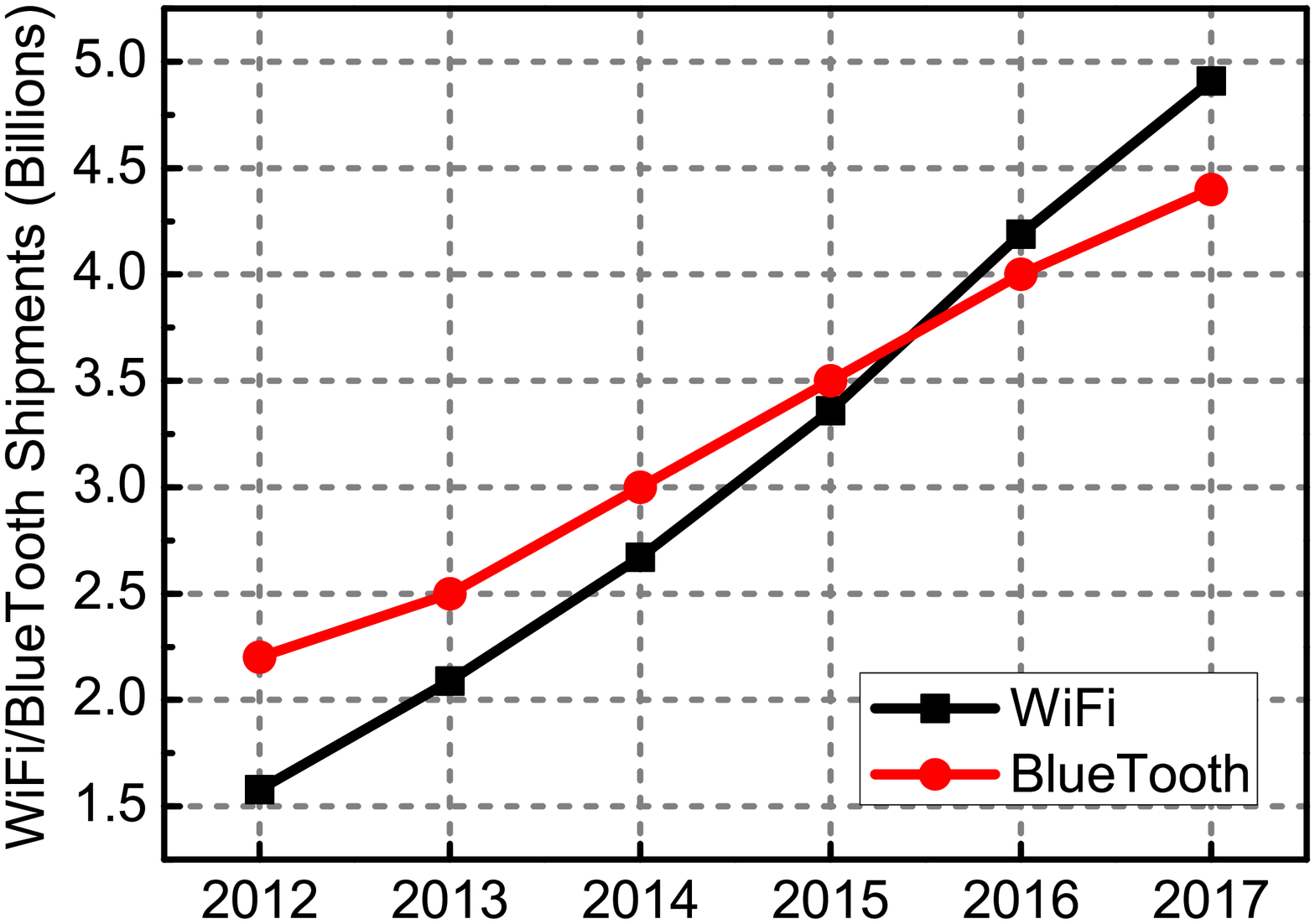}}
\subfigure[ZigBee Devices]{
\label{fig:zigbee}
\includegraphics[width=0.23\textwidth]{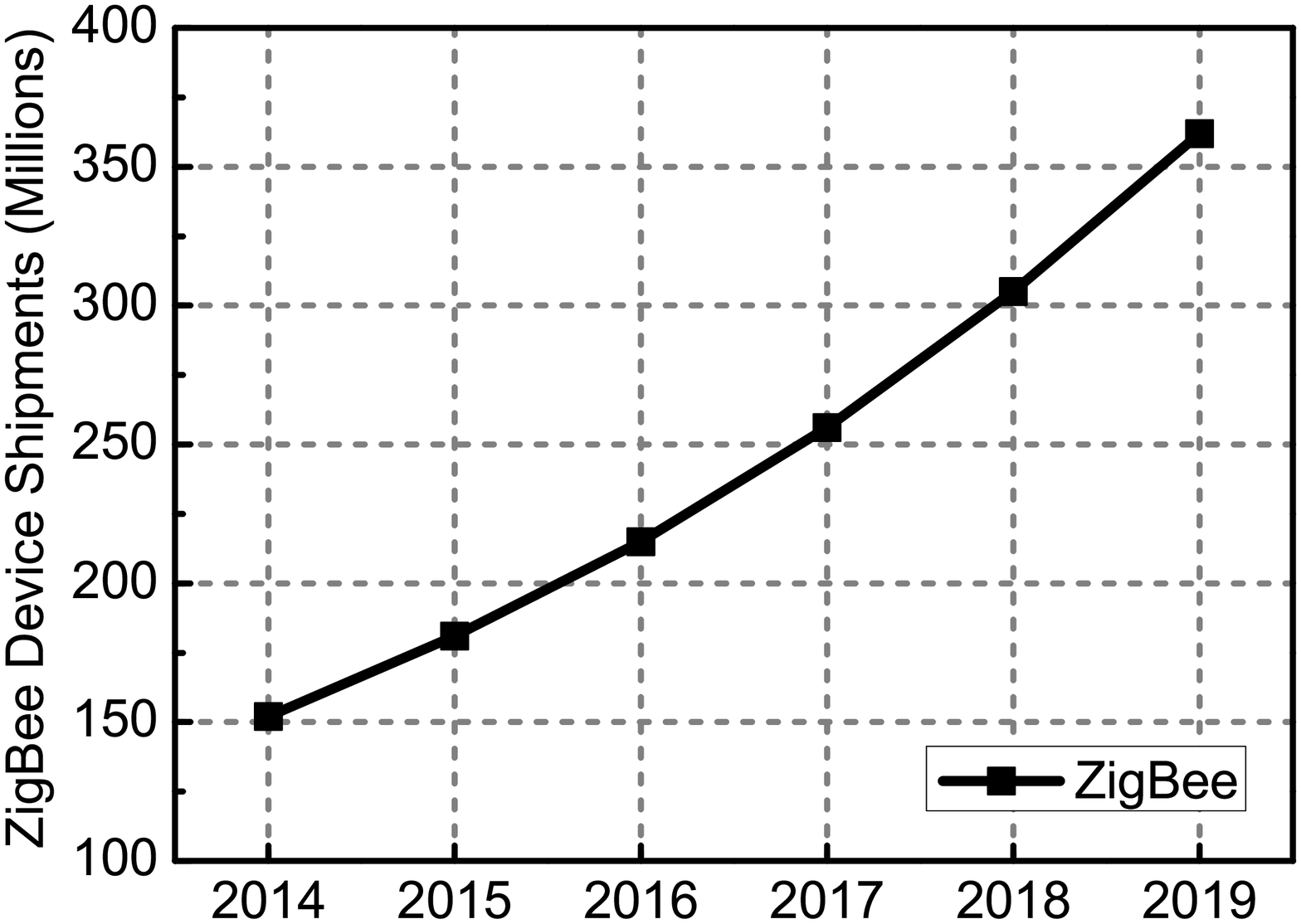}}
\caption{Annual WiFi, BlueTooth and ZigBee Device Shipments}
\label{wifibtzb}
\end{figure}

A wide range of wireless technologies, such as WiFi, BlueTooth and ZigBee
share the common wireless medium of the unlicensed 2.4GHz ISM band.
Figure~\ref{ctc:sharespectrum} shows the spectrum usage of IEEE 802.11b (WiFi), IEEE 802.15.1 (BlueTooth) and IEEE 802.15.4 (ZigBee),
from which we can see that many channels of these technologies are overlapped.
With the popularity of these technologies, the number of WiFi, BlueTooth, and ZigBee devices increases dramatically.
Figure~\ref{wifibtzb} plots the annual shipments of WiFi, BlueTooth and ZigBee devices.
As we can see, the shipments of WiFi, BlueTooth, and ZigBee have reached to 4.2 billions (Source: IC Insights), 4.0 billions (Source: IHS Technology), and 215 millions (Source: ABI Research and BI Intelligence Estimates) in 2016.

As so many devices put into use every year, the ISM band becomes more and more crowded and these heterogeneous devices
have to compete with each other to access the spectrum. Since existing wireless technologies lack the functionality of
harmonious and cooperative sharing the ISM band, they introduce cross-technology interference to each other which may cause
performance degradation or even destroy the wireless communication.

To address this issue, researchers propose cognitive radio to achieve dynamic spectrum access. While the cognitive radio
technology is proved to be effective in research, it encounters great resistance in real-world deployment since most of
existing devices have no cognitive capability and the cost of deploying commercial cognitive radio networks is extremely
high.
In the following of this paper, we focus on introducing spectrum efficient designs which are compatible
to existing standards and can be implemented directly on off-the-shelf devices.

\section{Related Work}\label{sec:Rela}

This section reviews related work on achieving spectrum efficient communication under cross-technology interference. 
%We focus on the studies which are compatible to existing standards and can be implemented directly on off-the-shelf devices.
We introduce related work on (i) link/channel modeling under CTI, and (ii) cross-technology
communication.

\subsection{Link Modeling under CTI}
Extensive link/channel models have been proposed to study the link/channel status.
These link/channel models include (i) hardware based models such as RSSI, LQI, and SNR,
and (ii) software based models including packet reception based models (e.g., PRR and KLE~\cite{KLE}),
require number of packet based models (e.g., RNP~\cite{Cerpa}, and Four-bit~\cite{fourbit}), and score
based models (e.g., WRE~\cite{XuWRE}, and F-LQE~\cite{Baccour}). These link models are proposed for
\emph{individual} links under intra-technology interference.

Our study shows that the characteristic of link/channel under
cross-technology interference is quite different from that
under intra-technology interference~\cite{CorLayer,cETX}. Cross-technology interference
introduces temporal and spatial dynamics to \emph{multiple}
adjacent links. This is because that (i) wireless communication
essentially occurs in a broadcast medium with concurrent receptions,
and (ii) the high-power heterogeneous devices (e.g., WiFi APs) causes
correlated loss on these links of low-power devices (e.g., ZigBee motes).
This new finding drives us to propose more realistic and accurate
link models~\cite{CorLayer,cETX} for the environment with cross-technology interference.

\subsection{Cross-Technology Communication}
In existing wireless communication environment, heterogeneous technologies (e.g., WiFi, BlueTooth, and ZigBee) co-exist in the same ISM band, which makes it possible to build a direct cross-technology communication (CTC) among these technologies.
Since these heterogeneous technologies have different corresponding PHY layers and MAC layers, one straightforward way is to either change the existing hardware or build new and dedicated hardware to enable the CTC. For example, the recent studies in~\cite{bharadia2015backfi,kellogg2015wi,zhang2013gap} require the design of additional low-power devices, which can directly send information to WiFi end devices.

However, the additional need of dedicated hardware makes cross-technology communication unrealistic in many applications. To enable the possible communication between billions of existing devices in the ISM band, researchers are focusing on building CTC between heterogeneous devices, while still following the existing hardware. The proposed approaches, Esense~\cite{chebrolu2009esense} and Howies~\cite{zhang2013howies}, enable the direct communication from commodity WiFi devices to commodity ZigBee devices by sending out dedicated WiFi packets. Since the ZigBee channel is overlapping with the WiFi channel, a ZigBee node is able to utilize the channel sensing technique (a mandatory for the CSMA purpose) to sense the channel energy. As a result, the ZigBee motes are able to recognize the different energy patterns in the ISM band, when there is and is not a WiFi packet.
To distinguish the energy patterns caused by the CTC packets and the background packets from other devices, these methods send out packets at different lengths and different rates, and further utilize various coding techniques.

Although these methods avoid the modification of the hardware, they do require sending out dedicated packets, thus lowering the channel efficiency. As a remedy for this, we propose CTC mechanisms~\cite{kim2015freebee,yin2017,jiang2017} rely on the existing traffic. For example, FreeBee~\cite{kim2015freebee} builds the CTC from WiFi to ZigBee via shifting the timings of the mandatory beacons without sending any additional packet. C-Morse~\cite{yin2017} and DCTC~\cite{jiang2017} enable the CTC between ZigBee and WiFi via the existing data traffic to further improve the CTC transmission rate.
In this way, these methods are channel efficient since they do not need to send out any additional packets, in contrast to Howies and Esense.

In addition, it is worth noting that multiple communication directions can be achieved in one strategy. For example, FreeBee achieves a bi-directional communication between WiFi and ZigBee, and single directional communication from BlueTooth Smart to WiFi and from BlueTooth Smart to ZigBee, while C-Morse and DCTC manage to provide a bi-directional between WiFi and ZigBee.

In summary, CTC aims at providing directing communication between heterogeneous devices with incompatible PHY layers. Our proposed CTC methods are able to achieve such a goal without hardware modification and extra traffic, and with better channel efficiency. 
\begin{figure}[!htbp]
\centering
\includegraphics[width=0.49\textwidth]{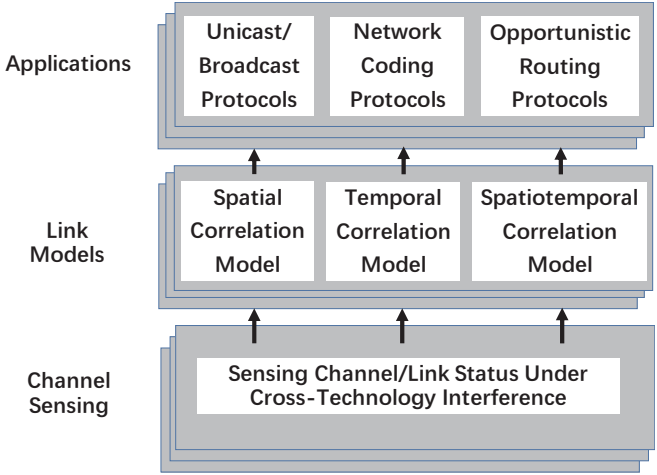}
\caption{The Overview of Link Modeling and Applications}
\label{linkmodel}
\end{figure}

\section{Link Modeling and Applications}\label{sec:Moti}

This section provides the overview of link modeling and its
applications under cross-technology interference, as shown
in Figure~\ref{linkmodel}. From the figure, we can see that
our designs sense the link/channel status under cross-technology
interference. We then summarize the characteristic of links/channels
under CTI and propose realistic link models such as spatial
link correlation model and temporal link correlation model.
Since our models accurately capture the link/channel status under
CTI, they have great potential to improve the performance
of upper layer designs such as unicast, broadcast, network coding
and opportunistic routing.

In the following of this section,
we first introduce the
impact of cross-technology interference to wireless links/channels.
Then, we introduce our \emph{framework} on link/channel models and the
potential applications with these models. The detailed technical
designs and evaluations will be omitted in the following description
but can be found in authors' publication~\cite{CorLayer,LCNCicnp,corlayerton,lcortwc,LCcause,unifiedlctwc,cETX}.

\subsection{The Impact of CTI}
We first introduce the impact of cross-technology interference to
wireless links/channels. In general, cross-technology interference
will introduce temporal and spatial dynamics on links/channels of
other heterogeneous technologies.

\noindent$\bullet$ \textbf{Temporal Dynamics:}
The interference pattern introduced by heterogeneous technologies are
quite different, depending on a technology's transmission rate, the length
of transmitted packets, adopted MAC schemes, as well as applications
(e.g., video streaming and web surfing).
For example, in data collection scenarios with the ZigBee technology,
the interference could be intermittent because of its sparse traffic.
Different from the ZigBee technology, WiFi may generate busty
interference when it deliver video streams. In summary, heterogeneous
technologies induce distinct temporal dynamics to the links/channels
of other technologies.

\noindent$\bullet$ \textbf{Spatial Dynamics:}
Wireless communication essentially occurs in a broadcast medium
with concurrent receptions. The transmissions of heterogeneous
devices (especially high-power devices) may cause packet reception
losses at multiple adjacent links of other heterogeneous devices
using the same channel. In this case, multiple nearby links may
lose same packet receptions simultaneously, and thus introducing
spatial dynamics to the links/channels.

\subsection{Link Modeling}
To capture the temporal and spatial dynamics introduced by cross-technology
interference, we introduce temporal and spatial link correlation models which
have great potential to improve the communication efficiency.

\noindent$\bullet$ \textbf{Temporal correlation model:}
In this model, temporal correlation represents the dependency among
consecutive transmissions that occurs within a short time duration.
The temporal correlation model helps us decide when should we transmit
packets. For example, busty interference may corrupt a series of
consecutive transmissions. With the help of temporal correlation model,
we can infer the success/failure of the following transmissions based
on the current transmission's reception information.

\noindent$\bullet$ \textbf{Spatial correlation model:}
In this model, spatial correlation represents the dependency among
multiple adjacent links departing from the same transmitter.
The spatial correlation model measures the spatial reception
diversity of multiple receivers. This spatial correlation information
is useful in designing protocols such as broadcast, network
coding, and opportunistic routing which essentially exploit the
diversity benefit of broadcast medium.

\subsection{Applications}
This section introduce the applications of our link models in unicast, broadcast, network coding,
and opportunistic routing.

\noindent$\bullet$ \textbf{Model Applications on Unicast:}
In unicast, it is meaningful to transmit packets when there is no cross-technology interference
(CTI) while avoiding useless transmissions
under CTI since the receiver can not successfully decode these transmissions. We thus propose temporary
link correlation model~\cite{cETX} which captures (i) temporal correlation between transmissions on a link
and (ii) spatiotemporal correlation between transmissions on adjacent multi-hop links.
Our temporary link correlation model is able to accurately predict whether following transmissions will
be successfully received or not based on the current status of packet receptions. Therefore, our model
helps existing unicast protocols (i) efficiently utilize unoccupied channels without CTI and (i) avoid
transmission failures under CTI. Our experiment results show that the transmission cost of various
state-of-the-art unicast protocols (e.g., OLSR~\cite{OLSR}, LQSR~\cite{LQSR} and srcRR~\cite{srcRR}) is reduced by
16\%$\sim$28\% with 0.7\% additional overhead.

\begin{figure}[!htbp]
\centering
\includegraphics[width=0.36\textwidth]{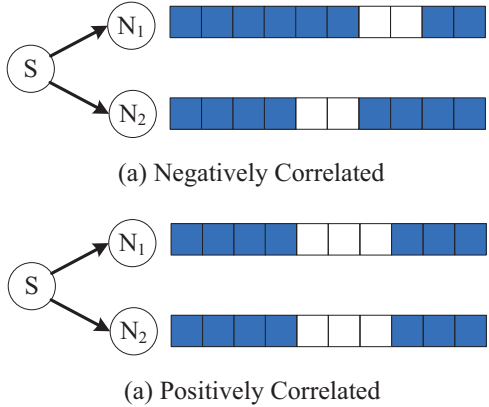}
\caption{The preference of broadcast on spatial correlation: (a) negative correlation and (b) positive correlation}
\label{ctc:broadcast}
\end{figure}

\noindent$\bullet$ \textbf{Model Applications on Broadcast:}
Broadcast, which delivers same content to multiple receivers simultaneously, is a fundamental operation in
wireless networks. Broadcast operations play a critical role in many network designs
such as routing discoveries and code dissemination. Under cross-technology interference, the broadcast packet receptions
at multiple receivers appear spatial dynamics, which inspires us to propose spatial correlation model
to capture this dynamics. 

For example, Figure~\ref{ctc:broadcast} shows negative
correlated (Figure~\ref{ctc:broadcast}(a)) and positive correlated (Figure~\ref{ctc:broadcast}(b)) packet reception
patterns at two receivers in a broadcast scenario. In this figure, blue blocks represent successful receptions
while white blocks represent packet losses at the two receivers.
In the negative-correlated packet reception scenario, we find that the link quality
from the source node $S$ to the two receivers $N_1$ and $N_2$ is 0.8 while the number of packets need to retransmit
is four. As a comparison, Figure~\ref{ctc:broadcast}(b) shows a positive correlated packet reception scenario where
we can see that the link quality is 0.7 and the number of packets need to retransmit is three. From this example,
we find that broadcast requires fewer number of retransmissions in positive correlated scenarios.

Based on this observation, the authors propose a transparent layer~\cite{CorLayer} which provides upper layer broadcast
protocols a logical topology and helps them form clusters with high positive correlation.
This design is integrated transparently with sixteen classic broadcast protocols on three
physical testbeds. The experimental results show that (i) our design significantly
improves the energy efficiency of these broadcast protocols and (ii)
the total number of broadcast transmissions is reduced by 47\% on average.

\begin{figure}[!htb]
\centering
\subfigure[Coding Scenario]{
\label{fig:BNZ}
\includegraphics[width=0.23\textwidth]{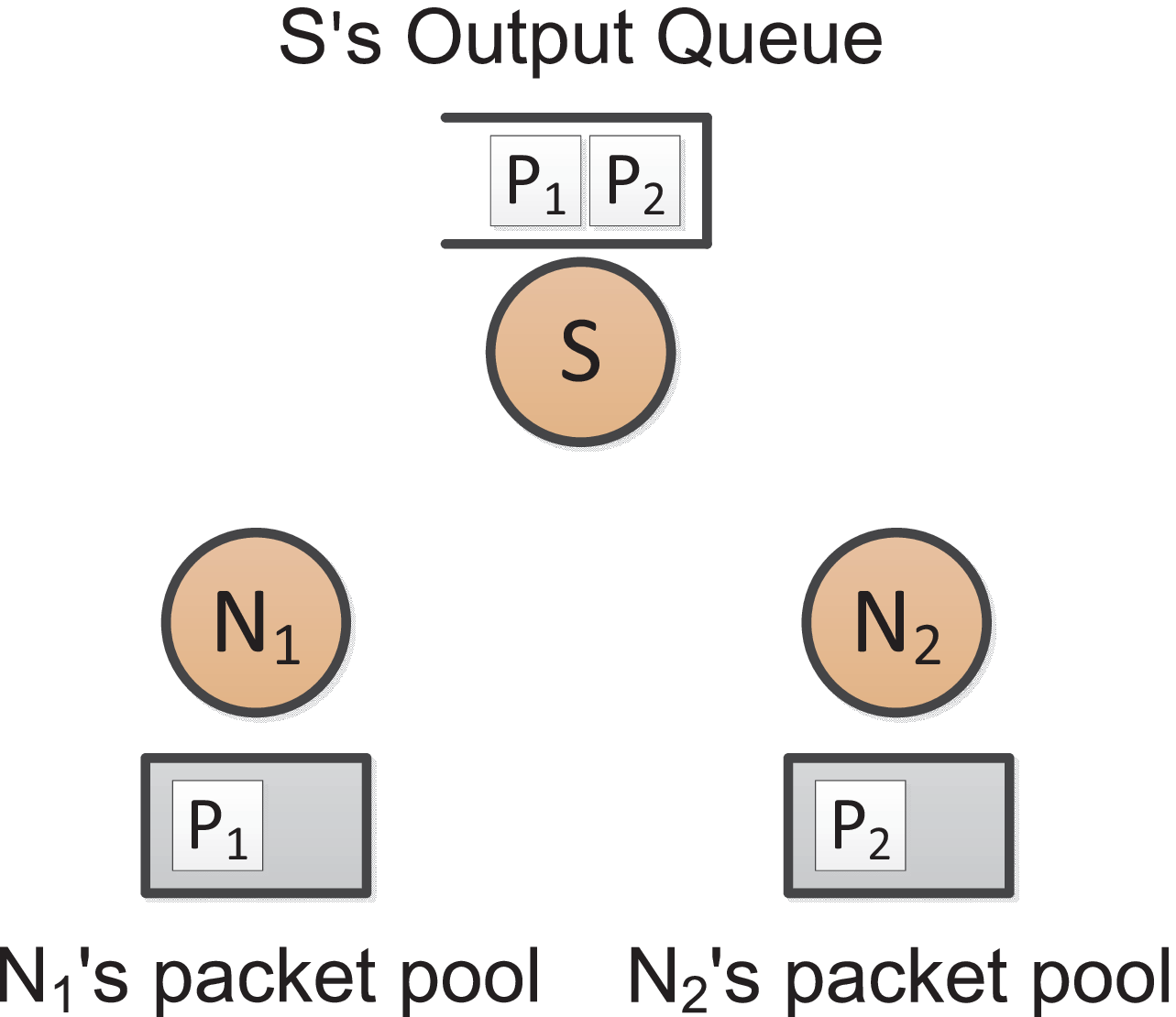}}
\subfigure[Non-coding scenario]{
\label{fig:UNZ}
\includegraphics[width=0.23\textwidth]{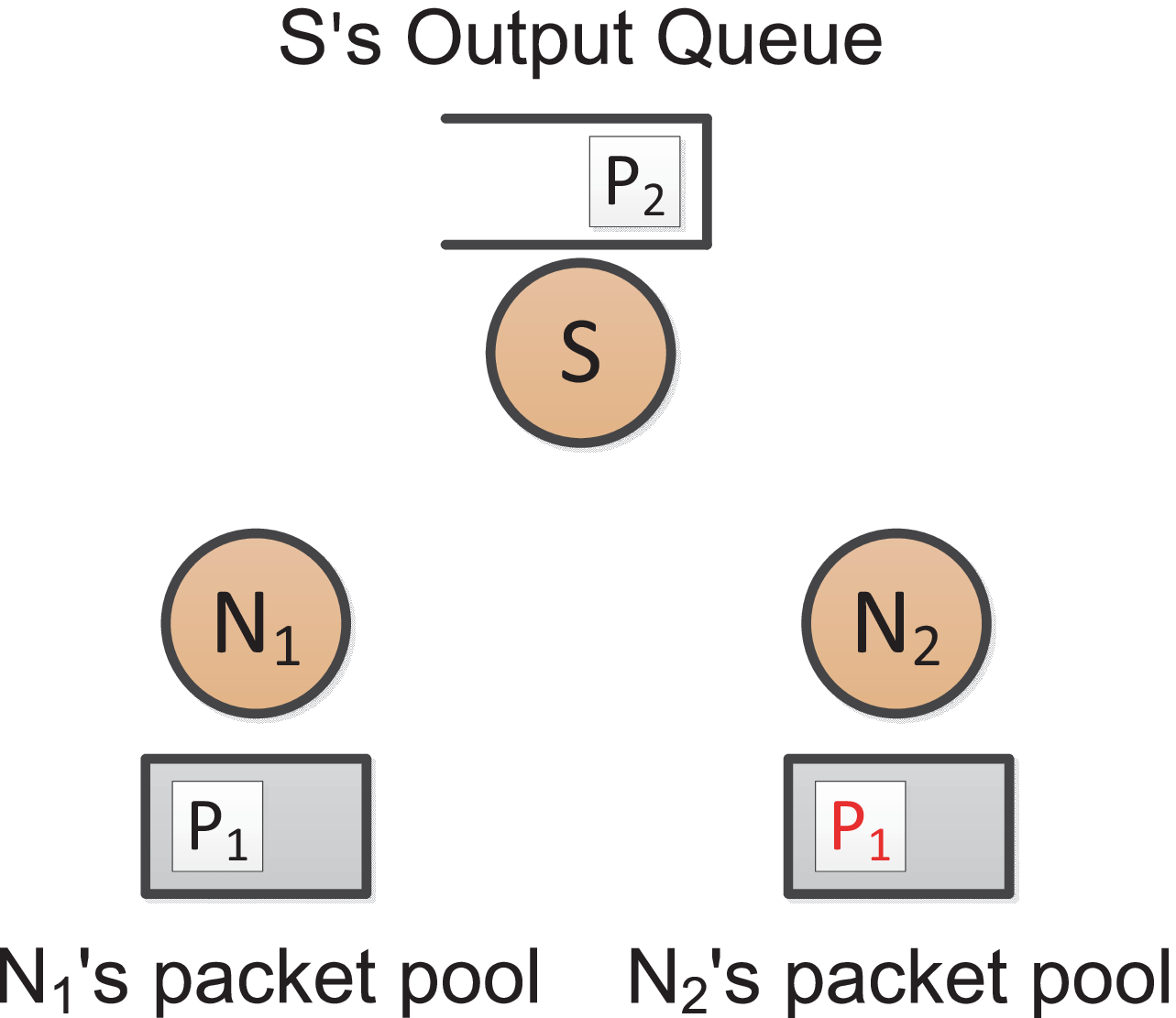}}
\caption{Network Coding: Coding Scenario vs. Non-coding scenario}
\label{networkcoding}
\end{figure}

\noindent$\bullet$ \textbf{Model Applications on Network Coding:}
The network coding technology allows the relay node to encode several packets together and send out with one transmission.
As shown in Figure~\ref{networkcoding}, there are two packets, i.e., $P_1$ and $P_2$ in sender $S$'s output queue.
Without network coding, the traditional design will send packet $P_1$ to receiver $N_2$ and send packet $P_2$ to receiver
$N_1$, using two transmissions. With network coding, the sender will send out a coded packet, i.e., $P_1+P_2$.
When the two receiver $N_1$ and $N_2$ receive the coded packet $P_1+P_2$, they will decode their desired packets through
minusing the packet in their packet pools from the coded packet.

The performance of network coding depends on each receiver's packet reception diversity.
For example, as shown in Figure~\ref{fig:UNZ}, if the receptions at the two receivers were
to be perfectly correlated (i.e., both receiver $N_1$ and $N_2$ has $P_1$ in their packet
pool), there are no coding opportunities at all and the total transmissions with and without
network coding would be the same. The performance of network coding may even worse than traditional
designs because of network coding's extra encoding/decoding overhead.

To address this issue,
authors propose a spatial correlation model~\cite{LCNCicnp} for network coding which helps
network designer/developers (i) decide whether they should apply the network coding technique given a
specific scenario and (ii) fully exploit the network coding benefits for those scenario which
are suitable for network coding.

\begin{figure}[!htbp]
\centering
\includegraphics[width=0.48\textwidth]{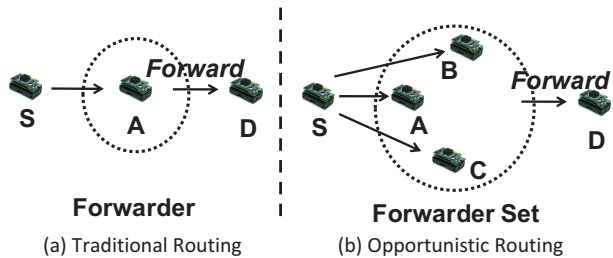}
\caption{Traditional Routing vs. Opportunistic Routing}
\label{ctc:or}
\end{figure}

\noindent$\bullet$ \textbf{Model Applications on Opportunistic Routing:}
Different from traditional routing, as shown in Figure~\ref{ctc:or}, a sender in opportunistic
routing is allowed to maintain a set of nodes as forwarders.
When the sender delivers packets to the next hop forwarders, once at least one node in
the forwarder set receives the packet, the forwarders with high priority will relay the
packet. From the working mechanism, we learn that opportunistic routing is exploiting
the spatial diversity, i.e., the reception diversity of multiple forwarder nodes.
Under cross-technology interference, it is highly possible that
all the forwarders lose a packet at the same time because of the high-power
interference. For example, the forwarder $A$, $B$, and $C$ in Figure~\ref{ctc:or}
may lose packets simultaneously. At this condition, opportunistic
routing not only degrades to traditional routing but also introduces extra
overhead in maintaining a large forwarder set (comparing with maintaining only one forwarder).

By applying spatial correlation model on opportunistic routing, the authors propose
correlation aware opportunistic routing~\cite{lcortwc} which is able to (i) fully exploit the
spatial diversity among nodes, (ii) help opportunistic routing select forwarder sets
with high spatial diversity, and (iii) avoid selecting useless nodes to the forwarder
set to reduce the overhead.

\begin{figure}[!htbp]
\centering
\includegraphics[width=0.49\textwidth]{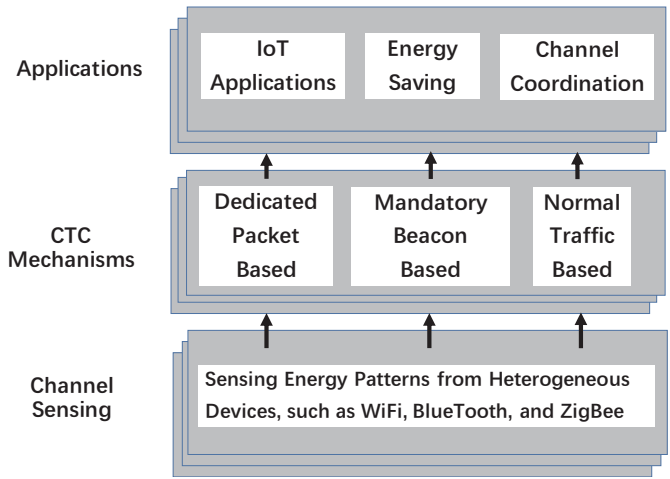}
\caption{The Overview of Cross-Technology Communication}
\label{ctcoverview}
\end{figure}

\section{Cross-Technology Communication}\label{sec:Test}

This section provides the overview of cross technology communication (CTC) mechanisms and its applications.
As shown in Figure~\ref{ctcoverview}, the CTC mechanisms sense energy patterns from heterogeneous devices
such as WiFi, BlueTooth, and ZigBee. Several different CTC mechanisms, including dedicated packet based, mandatory
beacon based, and normal traffic based approaches, are proposed.
These CTC mechanisms have broad applications. For example, the direct communication from WiFi to ZigBee can
be used to control sensors in smart home, thus bringing many new IoT applications. In addition,
CTC provides us a new to achieve channel coordination.

In the following of this section, we first introduce some background about cross-technology communication (CTC)
among heterogeneous devices and the possible benefits provided by CTC, followed by the possible ways to establish
the CTC. For the sake of clarity, we will focus on the framework of CTC. The technical details as well as evaluation
results can be found in authors' publication~\cite{kim2015freebee,yin2017,jiang2017}.

\subsection{CTC Background}

The introduced link model and the interference detection methods aim at providing a co-existence scheme between multiple heterogeneous devices while requiring no or little modification to the existing standard hardware. However, these methods are only implemented at the low-power receiver side, which will then lead to possible unfair issues among various types of heterogeneous devices. In this section, we will talk about cross-technology communication (CTC) which builds direct communication among heterogeneous devices via explicitly exchanging information. With the direct communication, heterogeneous devices are able to collaboratively work together to efficiently utilize the channel. In this way, CTC is able to achieve a goal which is similar to the link model and interference detection, but in a more fair manner. In addition, CTC enables additional benefits and applications given the rapid development of IoT, which will be talked about in the later part of this section.

\subsection{CTC Mechanisms}
In this section, we talk about the ways to enable the CTC between heterogeneous devices, which have different hardware constraints, different PHY layers and various MAC protocols.

The most straightforward way to build a direct link between heterogeneous devices is to design dedicated hardware~\cite{zhang2013gap,kellogg2015wi,bharadia2015backfi}. For example, In GSense~\cite{zhang2013gap}, the low-power ZigBee nodes first send out customized preambles to construct a special energy pattern in the air. By detecting this energy pattern, the WiFi device delays its transmissions to accommodate to ZigBee transmissions.  Although these methods are effective, they require either the modification of existing hardware or the deployment of new hardware, while the cost is too heavy for existing billions of commodity devices.

\begin{figure}[!htbp]
\centering
\includegraphics[width=0.4\textwidth]{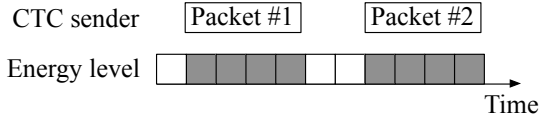}
\caption{An example of energy level change}
\label{ctc:energy}
\end{figure}

In contrast, recent CTC focus on build direct communication links, while still following the existing constraints, such as hardware assumptions, PHY and MAC layer protocols. Although the devices following different wireless technologies, such as WiFi and ZigBee, cannot directly decode the packets, the transmitted packets will lead to the change of the energy in the ISM band. As a result, these wireless devices are able to sample the channel energy by measuring the RSS (Received Signal Strength), which is a must for many commodity devices operating in the ISM band, such as WiFi, BlueTooth and ZigBee. Figure~\ref{ctc:energy} depicts the an example of the changing energy level when the sender transmits two packets.

However, the ISM band is crowded with wireless devices, all of which trying to compete the channel access. There will be inevitable wireless noises, which will affect the received energy patterns at the receiver side. Another challenge for enabling CTC is the RSS sampling speed at the low-power devices. For example, the ZigBee nodes, such as MICAz and TelosB motes, can only sample the SRRI at 31.25KHz, which might miss the transmitted CTC information.
To alleviate these issues, Esense~\cite{chebrolu2009esense} and Howies~\cite{zhang2013howies} choose to send out dedicated packets at specific lengths and specific rates, which will lead to special time durations different from the background noises. Coding techniques are further introduced to improve the reliability by combining the energy patterns of several packets to construct one CTC symbol.

\begin{figure}[!htbp]
\centering
\includegraphics[width=0.45\textwidth]{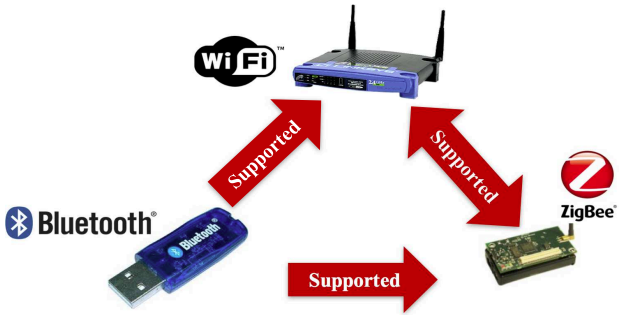}
\caption{The overview of FreeBee}
\label{ctc:freebee}
\end{figure}

As Esense and Howies rely on generating dedicated packets and the RSS sampling rate is inherently limited, these CTC methods will waste a lot of bandwidth for the CTC purpose. In contrast to these methods, we propose new CTC methods which
rely on existing opportunities to build the CTC. For example, FreeBee~\cite{kim2015freebee} constructs its CTC on the mandatory beacons, which usually have a period of 100ms. By shifting the transmission timings of these beacons, FreeBee utilizes the pulse position modulation (PPM) to construct its special energy patterns which will be recognized by utilizing the folding technique to filter out the noises.
Since the number of transmitted beacons is still the same, FreeBee does not incur any additional traffic overhead, meaning that it is a free side-channel. In additional to the bidirectional communication between WiFi and ZigBee, FreeBee also builds a unidirectional communication from BlueTooth to WiFi and from BlueTooth to ZigBee, as shown in Figure~\ref{ctc:freebee}.

In addition to the existing beacons, the data traffic in the dominating factor of the WiFi traffic. To boost the CTC throughput, we propose C-Morse~\cite{yin2017} and DCTC~\cite{jiang2017} for building the CTC based on a combination of WiFi data packets and WiFi beacons. Different from FreeBee, where beacons can be delayed arbitrarily, the data packets are delay sensitive and must be controlled very carefully. To meet the delay requirement of the data packets, both C-Morse and DCTC design special energy patterns, while at the same time maintaining the delay requirement of different applications. As a result, they can utilize the existing opportunities for constructing the CTC with a higher throughput, and also guarantee the legacy throughput. Similarly, $B^2W^2$~\cite{chi2016b2w2} proposed by Chi et.al builds a communication link from the WiFi to BlueTooth via the WiFi data traffic.

\subsection{CTC Applications}
To begin with, we introduce some representative applications by enabling CTC, such as (i) cost-efficient IoT applications, (ii) energy savings for power-hungry devices (e.g., WiFi), and (iii) channel coordination.

\noindent$\bullet$ \textbf{IoT Applications:}
In the deployment of current IoT systems, there are several different kinds of devices, each designed for the special purpose. For example, WiFi is introduced to provide a high speed connection to the internet, which can transmit the latest up-to-date information or the user's remote control. In contrast to WiFi, ZigBee is usually introduced to offer a relatively low transmission rate at the low-end and power constrained sensors, since the power consumption of ZigBee is much lower than that of WiFi. Because of these issues, the current deployed IoT systems require the exchange of information between heterogeneous devices. The most widely used method now is the deployment of multi-radio gateways a bridge for connecting them. For example, a WiFi and ZigBee dual radio gateway is able to receive the packet via 802.11 protocol, and then translates this message into corresponding ZigBee messages and then transmit them to the ZigBee motes, while the reverse direction can be done in the same way. Because of the requirement of the dual-radio gateway, these IoT systems need additional money, e.g. \$100 for one dual radio gateway, and the manual cost to deploy massive gateways for different devices from different product providers.

In contrast, CTC builds direct Device-to-Device (D2D) links between heterogeneous devices, without the requirement of the multiple-radio gateways. For example, CTC  is able to build a bi-directional communication between WiFi and ZigBee, while avoiding the change of hardware. In this way, CTC is able to provide a direct communication link between existing billions of WiFi and ZigBee devices with no hardware change and avoid the cost of deploying dual-radio gateways as well as the manual labor.

\begin{figure}[!htbp]
\centering
\includegraphics[width=0.49\textwidth]{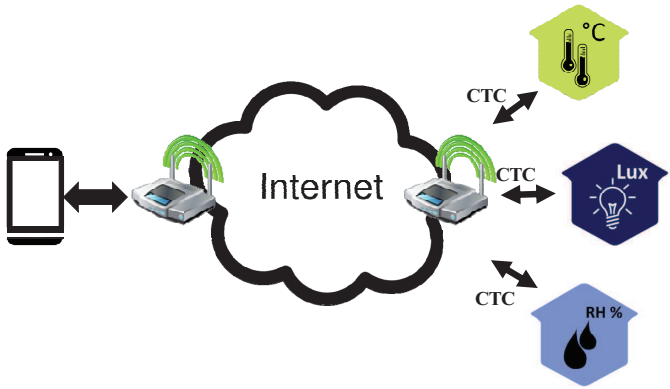}
\caption{IoT Applications via CTC}
\label{ctc:CTCapplication}
\end{figure}

One interesting application of CTC in IoT networks could be the efficient control of all sorts of sensors in smart-home scenarios. For example, people may deploy temperature, humidity, light and smoke sensors in their home. Those different sensors normally comes from different product providers and thus different gateways are provided. With the help of CTC, people can easily control various of sensors through their phones in anywhere without deploying any gateways. As shown in Figure~\ref{ctc:CTCapplication}, people can simply send their control instructions using their phones. This information is then delivered through Internet to the WiFi APs in their home. Finally, these WiFi APs can transmit the control information through the CTC technology to all kinds of sensors.

\begin{figure}[!htbp]
\centering
\includegraphics[width=0.49\textwidth]{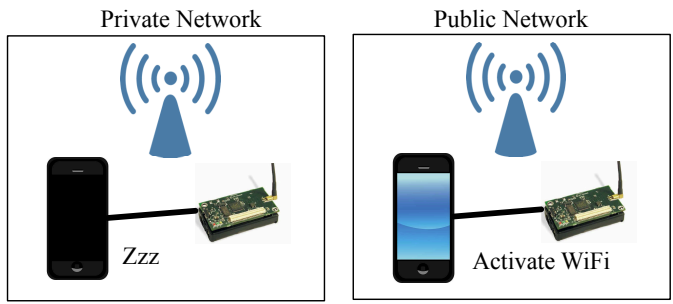}
\caption{Energy saving via CTC}
\label{ctc:energysaving}
\end{figure}

\noindent$\bullet$ \textbf{Energy Savings:}
In addition to the possible applications in IoT, CTC also manages to achieve a better energy efficiency.
Since the WiFi consumes much more energy that ZigBee, it is possible to save the energy of WiFi by adding one additional secondary ZigBee chip. Figure~\ref{ctc:energysaving} depicts an example of energy saving on the commodity smartphone which tries to connect to a
public AP. By utilizing the CTC, the WiFi chip is closed to save the energy consumption while the ZigBee chip is used to sense the public WiFi connections. The WiFi APs send information to ZigBee chips, which is similar to the periodical broadcasted WiFi beacons. The WiFi chips only wakes up when the ZigBee radio detects an available public WiFi AP. By reducing the time duration that the WiFi chips is working, CTC manages to achieve better energy efficiency for power-constrained devices, such as smartphones or laptops.

%\begin{figure}[!htbp]
%\centering
%\includegraphics[width=0.35\textwidth]{./figures1/CTC_channelallocation}
%\caption{Channel allocation via CTC}
%\label{ctc:ca}
%\end{figure}

\noindent$\bullet$ \textbf{Channel Coordination:}
In 802.11 protocol, WiFi devices utilize the RTS/CTS approach to explicitly announce the future channel status, so that other nodes will remain silent during this time period. Similar to this idea, CTC is able to build a
global RTS/CTS approach between heterogeneous devices. For example,
a WiFi device which has a huge volume of data traffic can broadcast the future channel status via CTC. When receiving this global CTS, ZigBee devices will delay transmission, since the low-power ZigBee packets will be corrupted by high power WiFi packets. By this
global RTS/CTS approach, various technologies can allocate the channel to avoid the interference between heterogeneous devices for better channel efficiency. Note that this idea can be also implemented on the ZigBee side, which can also use the global RTS/CTS to preempt the channel. When a CTC enabled WiFi device receive the global RTS/CTS, it will interpret this message via the 802.11 protocol, so that other devices which do not support CTC can also react as if they received the message from the ZigBee sensors.

\section{Conclusion}\label{sec:Conc}
This paper introduces our framework on achieving spectrum efficient communication
under cross-technology interference. The highlight of our work is that the proposed 
designs are compatible to existing standards and can be implemented directly on 
off-the-shelf devices. Our framework studies spectrum efficiency under cross-technology 
interference from two aspects. From the perspective of each communication technology
itself, we propose novel channel/link models to capture both temporal and spatial 
characteristics on channel/link status under cross-technology interference.
On the other hand, we propose cross-technology communication mechanisms for heterogeneous 
communication technologies which changes the spectrum competition problem to spectrum 
access negotiation through direct communication.  

\section*{Acknowledgement}
This work was supported in part by National Science
Foundation under grants CNS-1444021 and CNS-1525235.

\balance
\bibliographystyle{unsrt}
\bibliography{paper}

\end{document}